\documentclass[usenatbib, usegraphicx]{mn2e}

\newcommand{\dfrac}[2]{\displaystyle\frac{#1}{#2}}
\newcommand{\diff}[2]{\frac{\rmn{d} #1}{\rmn{d} #2}}
\newcommand{\pdiff}[2]{\frac{\partial #1}{\partial #2}}
\newcommand{\refp}[1]{(\ref{#1})}
\newcommand{\anglebr}[1]{\langle #1 \rangle}
\newcommand{\squarebr}[1]{\left[ #1 \right]}

\newcommand{\irm}{\rmn{i}}
\newcommand{\erm}{\rmn{e}}
\newcommand{\const}{\rmn{const}}
\newcommand{\pv}{\bmath{p}}

\newcommand{\tilrho}{\tilde\rho}
\newcommand{\tilPsi}{\tilde\Psi}
\newcommand{\tilTheta}{\tilde\Theta}
\newcommand{\tilvphi}{\tilde\varphi}
\newcommand{\tilG}{\tilde G}

\newcommand{\Pret}{P_\rmn{ret}}
\newcommand{\Pesc}{P_\rmn{esc}}
\newcommand{\Dq}{\Delta q}
\newcommand{\Dt}{\Delta t}
\newcommand{\Dqmean}{\anglebr{\Dq}}
\newcommand{\Dtmean}{\anglebr{\Dt}}
\newcommand{\sgmDq}{\sigma_{\Dq}}
\newcommand{\sgmDt}{\sigma_{\Dt}}
\newcommand{\PE}{P_\rmn{E}}
\newcommand{\PA}{P_\rmn{A}}

\newcommand{\tcycz}{\Dtmean_0}

\newcommand{\Vu}{V_\rmn{u}}
\newcommand{\Vd}{V_\rmn{d}}
\newcommand{\Vs}{V_\rmn{sh}}
\newcommand{\nuu}{\nu_\rmn{u}}
\newcommand{\nud}{\nu_\rmn{d}}
\newcommand{\Du}{D_\rmn{u}}
\newcommand{\Dd}{D_\rmn{d}}
\newcommand{\Vrel}{V_\rmn{rel}}
\newcommand{\Grel}{\Gamma_\rmn{rel}}

\newcommand{\Wsf}{W_\rmn{sf}}
\newcommand{\Wsfp}{W_\rmn{sf+}}
\newcommand{\Fsf}{F_\rmn{sf}}
\newcommand{\Fsfp}{F_\rmn{sf+}}
\newcommand{\Fsfpp}{F^{(p)}_\rmn{sf+}}
\newcommand{\Psip}{\Psi^{(p)}}
\newcommand{\Thetap}{\Theta^{(p)}}
\newcommand{\tilFsfp}{\tilde{F}_\rmn{sf+}}
\newcommand{\Ssfp}{S_\rmn{sf+}}

\newcommand{\mup}{\anglebr{\mu}_+}

\title[Probabilistic description of shock acceleration]
{
Probabilistic description of the first-order Fermi acceleration in shock waves:
Time-dependent solution by single-particle approach
}

\author[T.~N.~Kato and F.~Takahara]
{
Tsunehiko N. Kato
\thanks{\textit{Present address:}
Division of Theoretical Astrophysics,
National Astronomical Observatory,
2-21-1 Osawa, Mitaka, Tokyo 181-8588, Japan.
E-mail: tkato@th.nao.ac.jp}
and Fumio Takahara\\
Department of Earth and Space Science,
Graduate School of Science,
Osaka University,\\
Machikaneyama 1-1, Toyonaka, Osaka 560-0043, Japan
}

\date{in original form 27 March 2002, revised 6 February 2003}
\pubyear{2003}
\volume{000}
\pagerange{1-13}

\begin{document}
\maketitle

\begin{abstract}
We give a new coherent description
of the first-order Fermi acceleration of particles in shock waves
from the point of view of stochastic process of
the individual particles,
under the test particle approximation.
The time development of the particle distribution function
can be dealt with
by this description,
especially for relativistic shocks.
We formulate the acceleration process of a particle
as a two-dimensional Markov process
in a logarithmic momentum-time space,
and relate the solution of the Markov process
with the particle distribution function
at the shock front,
for both steady and time-dependent case.
For the case
where the probability density function of the energy gain and cycle-time
at each shock crossing of the particles obeys a scaling law in momentum,
which is usually assumed in the literature,
it is confirmed
in more general form
that
the energy distribution of particles
has the power-law feature in steady state.
The equation to determine the exact power-law index
which is applicable for any shock speed
is derived and it is shown that
the power-law index, in general, depends on
the shape of the probability density function of
the energy gain at each shock crossing;
in particular for relativistic shocks,
the dispersion of the energy gain
can influence the power-law index.
It is also shown that
the time-dependent solution has a self-similarity
for the same case.
\end{abstract}


\section{Introduction}
The first-order Fermi acceleration in shock waves
is a widely known mechanism
which generates non-thermal energetic particles
in space.
One of the most notable features
of this mechanism
is the power-law energy distribution of
accelerated particles in steady state.
The mechanism works in various shock waves
ranging
from the earth's bow-shock to
ultrarelativistic shocks
associated with gamma-ray bursts.
Near by the earth,
satellites and space-crafts
have directly observed
energetic particles accelerated by
this mechanism.
Recent X-ray observations
also discovered
synchrotron X-rays from energetic electrons
in several supernova remnants
(e.g. SN1006; see \citealt{Koyama95}).
These electrons
are considered to be accelerated in shock waves
by this mechanism
and have a power-law distribution.
These observations are believed to be evidence that
cosmic-rays with energies below $10^{15}$ eV (namely, `knee')
originate from supernova remnants in our galaxy.

Since the basic theory was proposed
in the late 1970's,
this mechanism has been investigated
by numerous authors
\citep[for review see][]{Drury83,Blandford87}.
While some of recent theoretical interests
are focused on
the non-linear problems \citep*{Drury81, EBJ96, Berezhko99},
some of the linear problems,
under the test particle approximation,
still remain to be clarified
and need more close examinations,
especially on
the time development of the particle distribution
and
the acceleration in relativistic shocks.
For non-relativistic shocks,
this mechanism is described well
by the diffusion-convection equation.
Solving this equation for steady state,
the power-law solution
is derived
\citep*{Axford77, Krymsky77, Blandford78}.
Time-dependent solutions
can also be examined
based on this equation,
usually with the aid of the Laplace transformation.
However,
to invert the transform in analytical form is 
generally difficult
and
analytical solutions were derived
only for several mathematically simple cases
\citep[e.g.][]{Toptyghin80}.
\citet{Drury91} proposed
an approximation for more general situations
by re-normalizing one of the analytical solutions
using the first two cumulants
of the particle distribution,
which can be obtained in analytical form.
\citet{Fritz90}
investigated
the effects of the synchrotron losses on
the time-dependent solution
for momentum-independent diffusion coefficients
and calculated solutions
by numerical inversion of the Laplace transforms.

Recently,
the particle acceleration in
relativistic shocks
has attracted some attention
in relation to AGN jets, GRBs
or ultra-high energy cosmic rays.
In this situation,
the problem becomes more difficult
because the anisotropy in the particle distribution
is not negligible and,
as a result,
the diffusion-convection equation is no longer valid.
Returning to the Boltzmann equation,
\citet{Kirk87a} derived the steady-state solution
by means of a semi-analytical approach,
performing the eigenfunction expansion.
This approach was recently extended to ultrarelativistic shocks
\citep{Kirk00}.
However,
the time development still has not been treated
in this way.

There is an alternative approach
established by \citet{Bell78}
which is based on the acceleration process
of individual particles.
From this approach,
the acceleration process is described as
\textit{a stochastic process} of individual particles;
particles are accelerated
whenever they repeat the cycle of crossing and re-crossing of the shock front,
where the energy gain per one-cycle and
the cycle-time are both regarded as stochastic variables.
The description needs not introduce the assumption
of the isotropy on the particle distribution.
It can therefore be naturally extended to
the acceleration in relativistic shocks
\citep{Peacock81}.
In our previous paper \citep{Kato01},
we investigated the acceleration process
in relativistic shocks
from this approach utilizing random walk theory.
However,
because
the previous studies dealt only with
the energy gain per one-cycle
and not with the cycle-time,
they are restricted within the steady problems
and can not treat the time development of
the distribution of particles,
even for non-relativistic shocks.
Furthermore,
the derivation of
the distribution function of particles
from this approach
and that of the power-law index
are still not satisfactorily established.

Monte Carlo simulations,
the third approach,
can make a direct estimation
of the distribution function
even for relativistic and ultrarelativistic shocks
\citep{Kirk87b, Ellison90, Ostrowski91, Bednarz96, Bednarz98}.
However,
it needs a large-scale simulation
to obtain sufficiently accurate results,
and physical interpretations of the results
in terms of analytical models
are also desirable.

In this paper,
we reanalyse the acceleration process
based on a stochastic, single-particle approach.
By treating the cycle-time as a stochastic variable explicitly,
our method can describe the time development of
the particle distribution function.
The method
is also applicable to relativistic
and ultrarelativistic shocks
as well as non-relativistic shocks.
The paper is organized as follows.
In Section \ref{sec:Description},
we formulate the mechanism
as a stochastic process.
In Section \ref{sec:Solution},
for a case where
the property of the one-cycle of the shock crossings
has a scaling law in momentum,
we investigate
steady and time-dependent solutions
of the distribution function of particles
by an analytical way.
In Section \ref{sec:Application},
as a check,
we apply this theory to non-relativistic shocks
and compare the results with conventional results.
In Section \ref{sec:Discussion},
we consider relations between the one-cycle properties
and the resulting particle distribution function
for both steady and time-dependent case.
Conclusions are given in Section \ref{sec:Conclusions}.

\section{Probabilistic description of the acceleration process}
\label{sec:Description}
In this section,
we formulate the acceleration process
of a particle
as a stochastic process,
and then relate it with the distribution
function of particles.
We also give numerical methods for obtaining
the solution.

\subsection{Basic acceleration process in shock waves}
From the single particle point of view,
the basic mechanism of the particle acceleration
is explained as follows.
First,
because of scattering caused by magnetic irregularity
existing in the plasma in both sides of the shock front,
particles move like random walk
and, as a result,
a fraction of the particles can repeat
crossing and re-crossing
of the shock front many times;
this can be modelled by that
in each cycle of the shock crossing,
a particle in the downstream region
returns to the shock front at a returning probability $\Pret$,
and otherwise escapes from the acceleration region.
Second,
since the electric field approximately vanishes
in the respective plasma (or `fluid') rest frames
in both regions,
the energy of a particle
measured in the respective rest frames
is unchanged
while the particle stays in one of the regions.
Consequently,
in a fixed reference frame,
particles gain energy
whenever they repeat the crossing cycle,
owing to the difference
between the fluid speeds;
in particular
for sufficiently relativistic particles,
letting the relative fluid speed between
the upstream and downstream region be $\Vrel (>0)$
and its Lorentz factor be $\Grel = (1-\Vrel^2/c^2)^{-1/2}$,
we can calculate the energy gain of a particle
for one-cycle of the shock crossings,
that is, a downstream to upstream to downstream cycle,
by performing two successive Lorentz transforms
together with the above condition:
\begin{equation}
	p_\rmn{f} = \Grel^2 \left( 1 - \frac{\Vrel}{c} \cos\theta_1 \right)
			\left( 1 + \frac{\Vrel}{c} \cos\theta'_2 \right) p_\rmn{i},
	\label{eq:energy gain}
\end{equation}
where $p_\rmn{i}$ and $p_\rmn{f}$
are the magnitudes of momentum of the particle
before and after the cycle, respectively.
$\theta_1$ and $\theta'_2$
are the angles at the shock crossing from the downstream to upstream region
measured in the downstream rest frame and
from the upstream to downstream region measured in the upstream rest frame,
respectively.
Considering the possible ranges of these angles to cross the shock front,
it is straightforwardly shown that
the energy always increases at every shock crossing cycle
(here, energy loss mechanisms are not taken into account).
Because
these angles are regarded as stochastic variables
on account of the scattering process,
the energy gain is also regarded as a stochastic variable.
Owing to the diffusive motion of the particles,
the cycle-time is also regarded as a stochastic variable.

\begin{figure}
  \includegraphics[width=\linewidth]{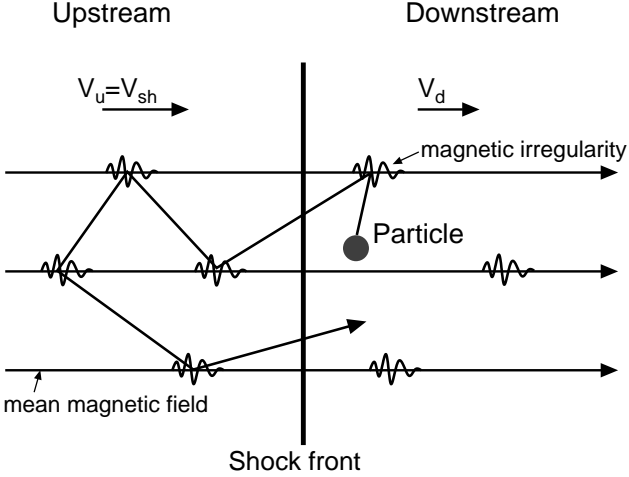}
  \caption{
	A schematic illustration of the acceleration of a particle
	 in a non-relativistic, parallel shock wave.}
  \label{fig:shock_accel}
\end{figure}

This is a basic picture of the acceleration mechanism
from the single particle point of view.
Fig.~\ref{fig:shock_accel}
illustrates the one-cycle of the acceleration process
of a particle
in a non-relativistic, parallel shock.

From a theoretical viewpoint,
the problem of the acceleration process consists of two parts.
One is how particles are accelerated
for \textit{one-cycle} of the shock crossings.
The other is
how the distribution function of the particles
evolves when the one-cycle properties are given.
The former is determined by details of the scattering process
and is quite a difficult problem
unless the diffusion approximation is applicable;
the properties of both turbulent magnetic field
in the vicinity of the shock front
and transport of particles in such turbulent fields
themselves remain long-standing problems.
On the other hand,
it is possible
to establish a well-defined theory formally
for the latter problem.
In the present paper,
we therefore construct that theory
from the point of view of the single particle approach.

\subsection{Description as a Markov process}
In this paper,
we investigate
the acceleration process
\textit{at the shock front};
our primary aim is to solve the steady state and time-dependent solution
of the distribution function of particles at the shock front.
Therefore,
we concentrate only on the state of particles, $(\pv,t)$,
at the crossing of the shock front,
where $\pv$ is the momentum of a particle and $t$ is the time respectively;
we measure $\pv$ in the downstream frame and $t$ in the shock rest frame,
respectively.
The history of acceleration process of a particle
observed at the shock front
can be described by
a sequence of the state of the particle
at each end of \textit{the acceleration cycle},
that is,
at each shock crossing from the upstream to downstream region:
\begin{equation}
	(\pv_0, t_0), (\pv_1, t_1), (\pv_2, t_2), ...\ .
\end{equation}
Such a sequence is terminated when
the particle escapes from the acceleration region,
usually from the downstream region.
We consider this process in the $\pv-t$ space
where
each state of the sequence defines a point,
which is called `state point' in the following.
If the changes in $\pv$ and $t$ at each cycle,
$(\Delta \pv, \Dt)$,
are stochastic variables
as already mentioned,
the acceleration process
can be regarded as a stochastic process in the $\pv-t$ space.
Usually,
this is a Markov process
because
the changes $(\Delta \pv, \Dt)$ mostly depend only
on the momentum at the last cycle.

Although
the history of $(\pv,t)$
provides complete information of
the acceleration process at the shock front,
we consider only the history of $(p,t)$ in this paper,
where $p$ is the magnitude of the momentum,
because
only $p$ concerns most practical purposes.
In order to treat the history of $(p,t)$ solely
as a Markov process,
we also assume that the probability density of the changes in $p$ and $t$
at one-cycle
does not depend on the direction of the momentum at the last cycle,
at least approximately.
(If this assumption is not appropriate,
we must deal with the state of a particle
by $(\pv, t)$ or $(p, \mu, t)$,
where $\mu$ is the cosine of the angle between
the shock normal and the momentum of the particle.)
For considering the acceleration of relativistic particles,
it is better to describe this process
in a logarithmic momentum space,
because the changes in $p$ at each cycle
are usually multiplicative rather than additive
as shown in equation \refp{eq:energy gain}.
Furthermore,
since we consider
only mono-energetic injection with a certain momentum of $p_0$
measured in the downstream rest frame
in the fundamental part of the following description,
we use a new quantity
\begin{equation}
	q := \ln (p/p_0)
	\label{eq:q def}
\end{equation}
instead of $p$
to simplify the description.

Thus,
the acceleration process of a particle can be considered as
a two-dimensional Markov process on the $q-t$ plane.
This stochastic process
is described
by a probability density function
which describes the transition on the $q-t$ plane for one cycle of the acceleration,
$\rho(\Dq, \Dt; q)$,
where $\Dq$ and $\Dt$ are respectively the changes in $q$ and $t$ at one-cycle;
$q$ in this notation denotes the logarithmic momentum \textit{at the last cycle}
representing the Markov property of this process.
(We also use the term `step' instead of `cycle'
for the point of view of Markov process in the following.)
Since a particle can be lost from the acceleration region
at each cycle at a probability of $\Pesc(q)$,
this density is a \textit{defective} one;
the integration of it over all area is \textit{not} normalized to unity,
but to the return probability, $\Pret(q) = 1-\Pesc(q)$, as follows:
\begin{equation}
	\int_0^\infty d\Dq \int_0^\infty d\Dt
	\  \rho(\Dq, \Dt; q) = \Pret(q).
\end{equation}
In the present paper,
for simplicity,
we assume $\Dq>0$,
that is,
any energy loss mechanisms
are not efficient.
(However,
to generalize the following description
to include the possibility of $\Dq < 0$
is not difficult.)
The correlation between the energy gain $\Dq$ and
the cycle-time $\Dt$,
which can be considerable for relativistic shocks
\citep[see][]{Bednarz96},
can be included in the functional form of $\rho$.
For later convenience,
we define here the moments of $\Dq$ and $\Dt$ as follows:
\begin{equation}
	\anglebr{(\Dq)^n}
	:= \frac{1}{\Pret} \int_0^\infty d\Dq \int_0^\infty d \Dt
			\ (\Dq)^n \ \rho(\Dq, \Dt; q),
\end{equation}
\begin{equation}
	\anglebr{(\Dt)^n}
	:= \frac{1}{\Pret} \int_0^\infty d\Dq \int_0^\infty d \Dt
			\ (\Dt)^n \ \rho(\Dq, \Dt; q),
\end{equation}
\begin{equation}
	\sgmDq := \sqrt{\anglebr{(\Dq)^2} - \Dqmean^2},
\end{equation}
\begin{equation}
	\sgmDt := \sqrt{\anglebr{(\Dt)^2} - \Dtmean^2}.
\end{equation}
These quantities are generally dependent on $q$ at the last cycle.
As already mentioned,
although
the functional form of $\rho(\Dq, \Dt; q)$
is determined by microscopic physics,
we do not specify it
and discuss the process formally in this paper
(we adopt only an approximate model of $\rho$ in Section \ref{sec:Application}
and simple toy-models in Section \ref{sec:Discussion}).

Defining the probability density function of the state points
at $n$th step,
$\rho_n$,
the progress
from $(n-1)$th step to $n$th step
is expressed by
\begin{eqnarray}
	\rho_n(q,t) = \int_0^q dq' \int_0^t dt' \ \rho(q-q', t-t'; q') \rho_{n-1}(q', t').
	\label{eq:rho_n}
\end{eqnarray}
Again,
the integrals of
these $\rho_n$'s over all area
are not normalized to unity,
but to 
the probability
at which
a particle continues the acceleration cycle \textit{at least} $n$ times, $P_n$:
\begin{equation}
	\int_0^\infty dq \int_0^\infty dt \ \rho_n(q,t) = P_n.
	\label{eq:rho_n normalization}
\end{equation}
Since
we consider now the case where
particles are all injected at the same state
of $(q_0,t_0) = (0, 0)$,
we have
\begin{eqnarray}
	\rho_1(q,t) = \rho(q,t; q=0).
\end{eqnarray}

The integral in equation \refp{eq:rho_n}
includes a plain convolution with respect to $t$.
It is better to represent such equations in the form of
the Laplace transform with respect to $t$.
Introducing the following notation
\begin{equation}
	\tilde{f}(s) := \int_0^\infty f(t) \erm^{-st} dt,
\end{equation}
where $f(t)$ is a function of $t$,
equation \refp{eq:rho_n} is transformed to
\begin{equation}
	\tilrho_n(q,s) = 
					\int_0^q \tilrho(q-q', s; q') \tilrho_{n-1}(q', s) \ dq'.
	\label{eq:rho_n s}
\end{equation}

In a mathematical sense,
the above stochastic process can be regarded as a terminating renewal process
in the probability theory
\citep[cf.][]{Feller,Cox}.
However,
since the present process is two-dimensional and also has a Markov property,
it is a more complex problem.
Furthermore,
the stochastic process defined by equation \refp{eq:rho_n s}
for a fixed $s$
can be regarded as
the random walk process
dealt with in the previous paper \citep{Kato01},
except that the present process is \textit{in $q$-space}
with the Markov property,
always increases the value of $q$ and has no absorbing barriers.

\subsection{The density of state points}
\begin{figure}
  \includegraphics[width=\linewidth]{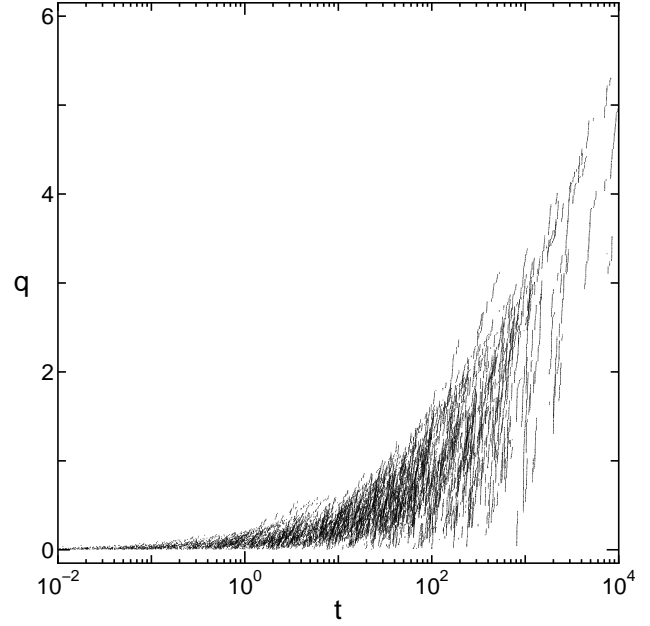}
  \caption{
  	A scatter plot of the state points
  	superposed for many particles
		obtained by Monte Carlo simulations
		(see Section \ref{sec:Application} for details).
		The unit of time is taken the mean cycle-time for $q=0$.
		The density of these points defines $\Psi(q,t)$
		in equation \refp{eq:Psi_sum}.
  }
  \label{fig:sequence}
\end{figure}

A history of acceleration process of a particle
can be represented by plotting the state points on the $q-t$ plane.
When such plots are made for many particles
and superposed as shown in Fig.~\ref{fig:sequence},
we can define the density of the state points:
\begin{equation}
	\Psi(q,t) := \sum_{n=1}^\infty \rho_n(q,t),
	\label{eq:Psi_sum}
\end{equation}
where the density is normalized by the number of the particles.
This is a key concept to treat the present problem.
The physical meaning of the product
$\Psi(q,t) dq dt$ is the expectation value of number of particles
which are initially injected at $(q,t)=(0,0)$
and then cross the shock front from the upstream to downstream side
within ranges $[q, q+dq]$ and $[t, t+dt]$,
per unit injection.
We can therefore relate this density with
the flux and
distribution function of the particles at the shock front
(see the following subsection).

Combining equations \refp{eq:rho_n} and \refp{eq:Psi_sum},
we obtain the following integral equation:
\begin{equation}
	\Psi(q,t) = \rho(q, t; 0) + \int_0^q dq' \int_0^t dt' \rho(q-q', t-t'; q') \Psi(q', t').
	\label{eq:Integral eq}
\end{equation}
This equation determines $\Psi$ from only $\rho$,
and is a fundamental equation to our description.
Taking the Laplace transform of this integral equation
with respect to $t$,
we obtain
\begin{equation}
	\tilPsi(q,s) = \tilrho(q, s; 0)
					+ \int_0^q \tilrho(q-q', s; q') \tilPsi(q', s) \ dq'.
	\label{eq:Integral eq s}
\end{equation}
For a fixed $s$,
this integral equation is regarded as
a Volterra equation of the second kind.
Since
it is generally difficult to solve this equation
in analytical form,
we will give numerical methods in Section \ref{sec:Numerical Solution}.
In the following,
we will use the following relation
\begin{equation}
	\int_0^\infty \Psi(q,t) dt = \tilPsi_0(q),
	\label{eq:Psi_t_inf-Psi0}
\end{equation}
where $\tilPsi_0(q) := \tilPsi(q,0)$.

Some quantities related with the acceleration process
are described in terms of $\Psi$ as follows.
From equations (\ref{eq:rho_n normalization}) and (\ref{eq:Psi_sum}),
we obtain
\begin{equation}
	\int_0^\infty dq \int_0^\infty dt \ \Psi(q,t)
	= \int_0^\infty \tilde\Psi_0(q) \ dq
	= \sum_{n=1}^\infty P_n = \langle n \rangle,
	\label{eq:ave n}
\end{equation}
where
$\langle n \rangle$ is the expectation value of the number of cycles
in an acceleration process.
(The probability at which
a particle continues the cycle {\it just} $n$ times
before escaping
is given by $P_n - P_{n+1}$.)
The probability at which
a particle escapes from the acceleration region
before attaining to $q$  for $0<t<\infty$
is given by
\begin{eqnarray}
	\PE(q)	&=& \Pesc(0) + \int_0^q dq' \Pesc(q') \int_0^\infty dt' \Psi(q',t')
					\nonumber \\
					&=& \Pesc(0) + \int_0^q \Pesc(q') \tilde\Psi_0(q') dq',
	\label{eq:PE}
\end{eqnarray}
because
a state point at $q$
does not make the next state point 
at the probability $\Pesc(q)$.
We can show $\PE(q) \to 1$ as $q \to \infty$.
The probability at which a particle
attains to $q$ for $0<t<\infty$
is of course given by $\PA(q) = 1 - \PE(q)$,
and we obtain from equation (\ref{eq:PE})
\begin{equation}
	-\diff{\PA(q)}{q} = \Pesc(q) \tilde\Psi_0(q).
	\label{eq:dPdp}
\end{equation}
Note that
$\PA(q)$ does not approach unity as $q \to 0$
but to $1-\Pesc(0)$
because $\Psi$ includes no contribution
from the state points of the particles just injected at $(0,0)$,
by its definition.
The mean acceleration time for attaining to $q$ is defined by
\begin{equation}
	\bar{t}(q) := \frac{\int_0^\infty t \Psi(q,t) dt}{\int_0^\infty \Psi(q,t) dt}
		= -\left. \pdiff{\tilPsi}{s}\right|_{s=0} / \tilPsi_0(q).
	\label{eq:accel time}
\end{equation}

From the point of view of the renewal process,
the density of state points $\Psi$ and
the integral equation \refp{eq:Integral eq} correspond to
the renewal density
and
the renewal equation,
respectively.
A concept similar to $\Psi$ was also introduced
in the previous paper \citep{Kato01}
to deal with the random walk of particles.

\subsection{Particle distribution function at the shock front}
\label{subsec:distribuion function}
Here,
we relate the density of state points introduced
in the previous subsection with
the particle distribution function at the shock front.
We consider
the acceleration of
relativistic particles in a plane shock,
where the velocity of the particles
can be approximated by the speed of light $c$
in the shock rest frame and in both fluid frames.

Consider first
an impulsive injection of unit particles
with $q=0$ at $t=0$
at the shock front.
Letting $\mu$ be the cosine of the angle between
the shock normal and the direction of particle momentum
measured in the shock rest frame,
we can define the distribution function of particles
at the shock front for this injection, $\Wsf(q,\mu,t)$,
where $q$ is measured in the downstream rest frame;
$\mu$ and $t$ are measured in the shock rest frame.
In terms of this function,
the one-sided flux of particles with logarithmic momentum $q$
at the shock front
from the upstream to the downstream
is represented by
\begin{equation}
	\Ssfp(q,t)	= \int_0^1 c \mu \ \Wsf(q,\mu,t) d\mu
						= c \mup \Wsfp(q,t),
\end{equation}
where we introduce the following notations:
\begin{equation}
	\Wsfp(q,t) := \int_0^1 \Wsf(q,\mu,t) d\mu,
\end{equation}
\begin{equation}
	\mup := \int_0^1 \mu \Wsf(q,\mu,t) d\mu \ / \ \Wsfp(q,t),
\end{equation}
where $\mup$ is, in general, dependent on $q$ .
On the other hand,
as already mentioned,
because
\begin{equation}
	\Psi(q,t) dt = \Ssfp(q,t) dt
\end{equation}
for the present situation,
we obtain the following relation
\begin{equation}
	\Wsfp(q,t) = \frac{1}{c \mup} \Psi(q,t).
	\label{eq:W-Psi relation}
\end{equation}
Because
this $\Wsfp(q,t)$ can be regarded as a Green's function
of the time-dependent solution of the distribution function
at the shock front,
the solution for more general injection rate $Q(t)$,
$\Fsfp(q,t)$,
is constructed as
\begin{eqnarray}
	\Fsfp(q,t)
	&=& \int_0^t \Wsfp(q,t-t') Q(t') dt'  \nonumber \\
	&=& \frac{1}{c \mup} \int_0^t \Psi(q,t-t') Q(t') dt',
\end{eqnarray}
where $\Fsfp$ is defined from $\Fsf$ like $\Wsfp$,
or in the form of the Laplace transform
\begin{equation}
	\tilFsfp(q,s) = \frac{1}{c \mup} \tilPsi(q,s) \tilde{Q}(s).
	\label{eq:Fs}
\end{equation}

In the following,
we deal with only the steady injection,
that is, $Q(t)=Q_0=\const$ for $t\ge0$
and $Q(t)=0$ for $t<0$.
For the case,
the steady state solution $(t \to \infty)$ is given by
\begin{equation}
	\Fsfp(q,\infty) = \frac{Q_0}{c \mup} \tilPsi_0(q),
	\label{eq:f0+}
\end{equation}
using the relation (\ref{eq:Psi_t_inf-Psi0}).
In the following,
to simplify the notations,
we also express the time-dependent solution
in terms of the `cut-off function' defined by
\begin{equation}
	\Theta(q,t) := \frac{\int_0^t \Psi(q,t') dt'}{\int_0^\infty \Psi(q,t') dt'}
							=  \frac{\int_0^t \Psi(q,t') dt'}{\tilPsi_0(q)},
	\label{eq:cut-off function}
\end{equation}
resulting in
\begin{equation}
	\Fsfp(q,t) = \Fsfp(q, \infty) \Theta(q,t).
	\label{eq:fzp-Theta}
\end{equation}
The Laplace transform of this function with respect to $t$
is written by
\begin{equation}
	\tilTheta(q,s) = \frac{1}{s} \frac{\tilPsi(q,s)}{\tilPsi_0(q)}.
	\label{eq:tilTheta}
\end{equation}
Because $\Psi(q,t)$ always takes non-negative value,
we can see
that $0 \le \Theta(q,t) \le 1$ for any values of $(q,t)$
for the steady injection;
the time-dependent solution always lies under
the envelope of the steady state solution.
This function is identical to
the function $\phi(t,x,p)$ of \citet{Drury91}
at the shock front $(x=0)$ [see equation (9) of his paper],
which was used
in the analysis of time-dependent solutions of the diffusion-convection equation.

It should be noted that,
in some cases,
the steady state solution
$\Fsfp(q,\infty)$ 
can be evaluated without introducing the density function $\Psi$.
Substituting equation \refp{eq:dPdp} into equation \refp{eq:f0+},
we obtain
\begin{equation}
	\Fsfp(q, \infty)
	= \frac{Q_0}{c \mup \Pesc(q)}
			\left( -\frac{d \PA(q)}{d q}\right).
\end{equation}
Therefore,
if $\PA(q)$ is estimated by another way,
the steady state solution is obtained immediately;
conventional single-particle approaches can estimate $\PA(q)$ approximately
\citep[e.g.][]{Bell78, Peacock81}
and therefore
can obtain the steady state solution.
Although,
to treat the time development of the distribution function,
we must treat the function $\Psi$ explicitly.

\subsection{Numerical Methods}
\label{sec:Numerical Solution}
By numerical methods,
the integral equation \refp{eq:Integral eq s}
can be solved straightforwardly
(see chapter 18.2 of \citealt{NR}).
For a given range $0<q<q_\rmn{max}$,
dividing the range into $N$ meshes
with uniform intervals of $h:=q_\rmn{max}/N$
and adopting the trapezoidal rule
to the integral in the equation,
and introducing notations
$q_i = i h$,
$\tilrho_{ij} := \tilrho(q_i-q_j,s;q_j)$
and $\tilPsi_i := \tilPsi(q_i,s)$
for $i=0,1,...,N$
(we omit the dependence on $s$ for simplicity here),
equation \refp{eq:Integral eq s} is discretized as follows:
\begin{equation}
	\tilPsi_0 = \tilrho_{00},
\end{equation}
\begin{equation}
	\left(1-\frac{h}{2} \tilrho_{ii} \right) \tilPsi_i
	= \tilrho_{i0}
	  + h \left( \frac{1}{2}\tilrho_{i0} \tilPsi_0
	       + \sum_{j=1}^{i-1} \tilrho_{ij} \tilPsi_j \right),
\end{equation}
where $i=1,2,...,N$.
Thus,
$\tilPsi_i$'s are trivially solved by forward substitution.
Note that the interval $h$ must be
fine enough to resolve the kernel $\tilrho$.

Once $\tilPsi(q,s)$ is obtained by this way for each $s$,
we can then make
a numerical inversion of Laplace transform
described in Appendix \ref{sec:Inversion}
to obtain $\Psi(q,t)$.
We can also obtain 
$\Fsfp(q,t)$ for general injection
or $\Theta(q,t)$ for steady injection
numerically
by inverting equation  \refp{eq:Fs} or \refp{eq:tilTheta}, respectively.

\section{Self-similar solution of particle distribution}
\label{sec:Solution}
One of the most well-known
features of the first-order Fermi acceleration
is the power-law distribution function
of accelerated particles
in steady state.
Theoretically,
it is just a consequent of some scaling laws,
or similarities,
in the one-cycle probability density function $\rho$;
in theoretical studies on the first-order Fermi acceleration,
such a scaling law is usually assumed implicitly or explicitly.
The scaling-law may realize
when, for example, the power spectrum of the turbulent magnetic field
has a power-law feature.
The fact that
the power-law distribution
is really observed in various shocks
may indicate such a scaling law
often realizes well.
In the following,
we also examine such a case
in an analytical way.

\subsection{One-step probability density function with a scaling law}
We consider here the case
in which
the transition of each step $(\Dq,\Dt)$ satisfies
the following scaling law;
for normalized variables 
$\Dq$ and $\Dt/(p/p_0)^\alpha (= \erm^{-\alpha q} \Dt)$,
where $\alpha$ is a constant parameter,
the probability distribution function defined for these variables
is independent of $q$ at the last step
or, in other words,
common for all $q$.
This means that
the one-step probability density $\rho(\Dq, \Dt; q)$
can be represented by a single function $\varphi(\Dq, \erm^{-\alpha q} \Dt)$ as
\begin{equation}
	\rho(\Dq, \Dt; q)
	= \erm^{-\alpha q} \varphi ( \Dq, \erm^{-\alpha q} \Dt )
	\label{eq:rho_sym}
\end{equation}
and
\begin{equation}
	\tilrho(\Dq, s; q) = \tilvphi \left( \Dq, e^{\alpha q} s \right).
	\label{eq:phi}
\end{equation}
The similarity of this function is clear.
Note that
$\Pret(q)$, $\Pesc(q)$ and
the moments of $\Dq$
are all independent of $q$:
\begin{equation}
	\Pret = \int_0^\infty \tilvphi_0(\Dq) d\Dq = \const
\end{equation}
and
\begin{equation}
	\anglebr{(\Dq)^n}
	= \frac{1}{\Pret} \int_0^\infty (\Dq)^n \tilvphi_0(\Dq) d\Dq
	= \const,
\end{equation}
where $\tilvphi_0(\Dq) := \tilvphi(\Dq, s=0)$.
The integral equation (\ref{eq:Integral eq s}) is reduced to
\begin{equation}
	\tilPsi(q,s) = \tilvphi( q, s )
							+ \int_0^q
							\tilvphi( q-q', \erm^{\alpha q'} s ) \tilPsi(q', s) \ dq'.
	\label{eq:Integral eq2}
\end{equation}
%

\subsection{Steady state solution}
\subsubsection{Power-law solution}
We first consider the steady state solution here.
Setting $s=0$ in equation \refp{eq:Integral eq2},
we have
\begin{equation}
	\tilPsi_0(q) = \tilvphi_0( q )
							+ \int_0^q \tilvphi_0( q-q' ) \tilPsi_0(q') \ dq'.
	\label{eq:Integral eq s0}
\end{equation}
Again,
denoting the Laplace transform
with respect to $q$ by
\begin{equation}
	f^*(\theta) := \int_0^\infty f(q) e^{-\theta q} dq,
\end{equation}
where $f(q)$ is a function of $q$,
we can write the exact solution of equation \refp{eq:Integral eq s0}
in the form of the Laplace transform as follows
\begin{equation}
	\tilPsi^*_0(\theta) = \frac{\tilvphi^*_0(\theta)}{1 - \tilvphi^*_0(\theta)}.
\end{equation}
As well known,
the inversion of this transform is given through
the Bromwich integral on the complex $\theta$-plane:
\begin{equation}
	\tilPsi_0(q)
		= \frac{1}{2 \pi \irm} \int_{\gamma-\irm\infty}^{\gamma+\irm\infty}
					\frac{\tilvphi^*_0(\theta)}{1 - \tilvphi^*_0(\theta)} \erm^{\theta q} d\theta,
	\label{eq:Bromwich}
\end{equation}
where $\gamma$ is a real constant
taken so that the vertical line $\rmn{Re}(\theta)=\gamma$
lies on the right of the all poles of
$\tilvphi^*_0(\theta) / [1 - \tilvphi^*_0(\theta)]$.
While the numerical solution
can be obtained directly (see Appendix \ref{sec:Inversion}),
we investigate
the asymptotic behaviour of equation \refp{eq:Bromwich} here.

Assuming that $\tilvphi^*_0(\theta) \to 0$ as $|\theta| \to \infty$
(we expect this condition usually holds),
the contour of the integral in the last equation
is closed through the left of the $\theta$-plane.
The integrand obviously has poles
at the points where the condition $\tilvphi^*_0(\theta) = 1$
is satisfied.
Because
$\tilvphi^*_0(0) = \Pret <1$ and
\begin{equation}
	\diff{\tilvphi^*_0}{\theta}
	= -\int_0^\infty \Dq \tilvphi_0(\Dq) e^{-\theta \Dq} d\Dq <0,
\end{equation}
the equation
$\tilvphi^*_0(\theta) = 1$ has the unique negative root
on the real axis of the complex $\theta$-plane.
We write this root as $\theta_-$.
As $q$ becomes sufficiently large,
because only the pole at $\theta_-$ dominantly contributes to
the integral \refp{eq:Bromwich},
we obtain the asymptotic solution $(q \to \infty)$
\begin{equation}
	\tilPsi_0(q) \sim A^{-1} \erm^{-\lambda q},
	\label{eq:Psi0 asymptotic}
\end{equation}
where
$\lambda := -\theta_-$ ($\lambda>0$)
and
\begin{equation}
	A := - \left. \diff{\tilvphi^*_0}{\theta}\right|_{\theta=-\lambda}
		= \int_0^\infty \Dq \tilvphi_0(\Dq) \erm^{\lambda \Dq} d\Dq.
	\label{eq:A}
\end{equation}
The index $\lambda$ is, by definition of $\theta_-$,
determined as the unique positive root of
\begin{equation}
	\int_0^\infty \tilvphi_0(\Dq) \erm^{\lambda \Dq} d\Dq = 1.
	\label{eq:lambda determine}
\end{equation}
Recalling $q=\ln(p/p_0)$,
the steady state solution (\ref{eq:f0+}) for large $q$
is written by
\begin{equation}
	\Fsfpp(p,\infty)
		\sim \frac{Q_0}{A c \mup}
			\frac{1}{p_0} \left( \frac{p}{p_0} \right)^{-(1+\lambda)},
	\label{eq:fzp asymptotic}
\end{equation}
where $\Fsfpp(p) dp = \Fsfp(q=\ln(p/p_0)) dq$.
This is the power-law solution
whose index is given by $\sigma = 1 + \lambda$.

\subsubsection{Approximations}
\label{sec:approximation steady}
If the dispersion of $\Dq$, $\sgmDq$, is negligible
in equations (\ref{eq:lambda determine}) and (\ref{eq:A}),
namely $\sgmDq \ll \lambda^{-1}$,
we can approximate
$\tilvphi_0(q) = \Pret \delta(q-\Dqmean)$ there,
resulting in
\begin{equation}
	\lambda = -\frac{\ln \Pret}{\Dqmean}.
	\label{eq:lambda approx1}
\end{equation}
This expression is equivalent to equation (5) of \citet{Peacock81}.
We will mention the validity of this formula again
in Section~\ref{sec:Discussion}.
The condition to use this approximation
can be rewritten as
\begin{equation}
	\frac{\Dqmean}{\sgmDq} \gg -\ln\Pret.
	\label{eq:approx_criterion}
\end{equation}
Furthermore,
if $\sgmDq \ll \Dqmean$,
we have
\begin{equation}
	A = \Dqmean.
\end{equation}
On the other hand,
if $\Dqmean \ll \lambda^{-1}$ and $\sgmDq \ll \lambda^{-1}$,
expanding $\exp(\lambda q)$
to the first-order of $\lambda q$
in both equations (\ref{eq:lambda determine}) and (\ref{eq:A}),
we obtain
\begin{equation}
	\lambda = \frac{\Pesc}{\Pret \anglebr{\Dq}},
	\qquad
	A = \Pret \anglebr{\Dq}
	\label{eq:lambda A approx2}.
\end{equation}
As shown in Section \ref{sec:Application},
this approximation is applicable
to the acceleration in non-relativistic shocks.

\subsection{Time-dependent solution}
\subsubsection{Self-similarity}
Here,
we show that
the time-dependent solution of equation \refp{eq:Integral eq2}
has a self-similarity for $q \gg \Dqmean$.
Firstly,
for $q \gg \Dqmean$,
we have approximately
\begin{equation}
	\tilPsi(q,s) = \int_0^q \tilvphi(q-q', e^{\alpha q'} s) \tilPsi(q',s) dq'.
	\label{eq:relation asym}
\end{equation}
This equation must be regarded as a \textit{relation}
for large $q$,
\textit{not} as an integral equation defined over all region of $q$;
such an integral equation would have only a trivial zero solution.
For a positive constant $a$,
it is seen that
the function defined by
\begin{equation}
	\tilPsi_a(q,s) := \tilPsi(q+\ln a, a^{-\alpha}s)
\end{equation}
also satisfies the above relation instead of $\tilPsi$;
in other word,
the relation does not change its form
under the transformation $q \to q + \ln a$ and $s \to a^{-\alpha}s$
simultaneously.
Secondly,
the steady state solution \refp{eq:Psi0 asymptotic},
which is regarded as a `boundary condition' at $s=0$,
is only multiplied by the constant $a^{-\lambda}$
under this transformation.
Finally,
because the relation \refp{eq:relation asym} is linear,
we obtain the following \textit{self-similarity}
of the density of state points for $q \gg \Dqmean$:
\begin{equation}
	\tilPsi(q+\ln a, a^{-\alpha} s) = a^{-\lambda} \tilPsi(q,s),
	\label{eq:self-similar_tilPsi}
\end{equation}
\begin{equation}
	\Psi(q+\ln a, a^{\alpha} t) = a^{-(\alpha+\lambda)} \Psi(q,t).
	\label{eq:self-similar_Psi}
\end{equation}
For the cut-off function,
we obtain
\begin{equation}
	\Theta(q+\ln a, a^\alpha t) = \Theta(q,t).
\end{equation}
Defining functions for $p$ instead of $q$,
$\Psip(p) dp := \Psi(q=\ln(p/p_0)) dq$
and $\Thetap(p) := \Theta(q=\ln(p/p_0))$,
we obtain
\begin{equation}
	\Psip(a p, a^\alpha t) = a^{-(1 + \alpha + \lambda)} \Psip(p,t),
\end{equation}
\begin{equation}
	\Thetap(a p, a^\alpha t) = \Thetap(p,t).
\end{equation}

For later convenience,
we derive a relation between
the steady state solution
and the time-dependent solution here.
Because $\varphi(q,t)$ is
usually a concentrated function of $t$,
the Laplace transform of it, $\tilvphi(q,s)$,
has a characteristic value of $s=s_*$
so that $\tilvphi(q,s)$ can be approximated
by $\tilvphi_0(q)$ for $s < s_*$.
Therefore,
in the region of $q-s$ plane where the condition
$\erm^{\alpha q}s < s_*$ is satisfied, 
the integral equation \refp{eq:Integral eq2}
is approximated by
\begin{equation}
	\tilPsi(q,s)
		= \tilvphi_0(q) + \int_0^q \tilvphi_0(q-q') \tilPsi(q',s) dq',
\end{equation}
resulting in $\tilPsi(q,s) \sim \tilPsi_0(q)$ in that region.
Thus,
we can define
\begin{equation}
	q_*(s) := \frac{1}{\alpha} \ln(s_*/s)
	\label{eq:q_ast}
\end{equation}
so that
we have $\tilPsi(q,s) \sim \tilPsi_0(q)$
for $s<s_*$ and $q<q_*(s)$.

\subsubsection{Approximate solution}
\label{sec:approximation time dep}
When
$\tilPsi$ varies with $q$ much more slowly than $\tilvphi$,
we can approximate
in the integrand of equation (\ref{eq:Integral eq2}) as
\begin{equation}
	  \tilPsi(q',s) \sim \tilPsi(q,s) + (q'-q)\pdiff{\tilPsi}{q}(q,s).
\end{equation}
A rough criterion to use this approximation
may be given by $\Dqmean \ll \lambda^{-1}$
and $\sgmDq \ll \lambda^{-1}$,
since the typical variation scale of $\Psi$ is given by $\lambda^{-1}$
for the steady state solution \refp{eq:Psi0 asymptotic}.
Thus,
for $q \gg \Dqmean$,
the integral equation \refp{eq:Integral eq2} is reduced to
\begin{equation}
	\tilPsi(q,s) = B_1(q,s) \tilPsi(q,s) - B_2(q,s) \pdiff{\tilPsi}{q}(q,s),
	\label{eq:B_differential}
\end{equation}
where
\begin{eqnarray}
	B_1(q,s) &:=& \int_0^\infty \tilvphi(\Dq, \erm^{\alpha q} s) d\Dq,
	\label{eq:B1}\\
	B_2(q,s) &:=& \int_0^\infty \Dq \tilvphi(\Dq, \erm^{\alpha q} s) d\Dq.
	\label{eq:B2}
\end{eqnarray}
For a fixed $s$,
equation \refp{eq:B_differential}
can be regarded as an ordinary differential equation for $\tilde\Psi$,
and the solution is given by
\begin{equation}
	\tilPsi(q,s) =
	C(s)\exp\left[ -\int_{q_0(s)}^q \frac{1-B_1(q',s)}{B_2(q',s)} dq'\right],
	\label{eq:tilPsi_solution_1}
\end{equation}
where $C(s)$ and $q_0(s)$ are functions of $s$;
we can choose the function $q_0(s)$ arbitrarily here.
If we choose $q_0(s) = q_*(s)$,
which was defined in equation \refp{eq:q_ast},
we can determine $C(s) = \tilPsi_0(q_*(s))$,
because $\tilPsi(q,s) = \tilPsi_0(q)$ for $q \le q_*(s)$.
Thus,
the solution is written by
\begin{equation}
	\tilPsi(q,s) =
		\tilPsi_0(q_*(s)) \exp\left[ -\int_{q_*(s)}^q \frac{1-B_1(q',s)}{B_2(q',s)} dq'\right].
\end{equation}
It is easily seen that
the above expression satisfies
the similarity \refp{eq:self-similar_tilPsi},
because $B_1(q+\ln a, a^{-\alpha} s) = B_1(q,s)$,
$B_2(q+\ln a, a^{-\alpha} s) = B_2(q,s)$
and $q_*(a^{-\alpha} s) = q_*(s) + \ln a $.
For $q<q_*(s)$,
because
$B_1(q,s) = \Pret$ and $B_2(q,s) = \Pret \Dqmean$,
we obtain
\begin{equation}
	\int_{q_*(s)}^q \frac{1-B_1(q',s)}{B_2(q',s)} dq'
	= \frac{\Pesc}{\Pret \Dqmean} (q-q_*(s)),
\end{equation}
which is consistent with the steady state solution \refp{eq:Psi0 asymptotic}
with $\lambda$ of equation \refp{eq:lambda A approx2}.
For $q \gg \Dqmean$,
where
the asymptotic solution \refp{eq:Psi0 asymptotic} is applicable
for steady state,
we finally obtain the following asymptotic expression
\begin{equation}
	\tilPsi(q,s) = A^{-1} \erm^{-I(q,s)},
	\label{eq:approx_tilPsi}
\end{equation}
where
\begin{equation}
	I(q,s) := \int_0^q \frac{1-B_1(q',s)}{B_2(q',s)} dq'.
\end{equation}
We will use this approximation
in the following section.
Although
it is usually difficult
to invert the above expressions
analytically,
it is able to do it
numerically (see Appendix \ref{sec:Inversion}).
Under the above approximation,
the mean acceleration time defined by \refp{eq:accel time}
is represented by
\begin{equation}
	\bar{t}(q) = \frac{\erm^{\alpha q} -1 }{\alpha \Pret \Dqmean} \tcycz,
\end{equation}
where $\tcycz$ denotes the mean cycle-time $\Dtmean$ for $q=0$.

\section{Application to non-relativistic shocks}
\label{sec:Application}
In order to confirm that
our method surely reproduces
the well-known results,
we apply the method to the acceleration in non-relativistic shocks
in this section.
We consider a simple one-dimensional parallel shock
as in Fig.~\ref{fig:shock_accel}
with shock speed $\Vs$($\ll c$).
The fluid speeds in the upstream and downstream region
measured in the shock rest frame
are denoted by $\Vu$ and $\Vd$, respectively,
and are assumed uniform in each region.
The compression ratio is given by $r=\Vu/\Vd$.

\subsection{One-step probability density function}
When we apply our method to a specific case,
we must first specify the one-step probability density $\rho$.
Assuming that $\rho$ obeys the scaling law \refp{eq:rho_sym}
discussed in the previous section,
we derive here an approximate expression of $\varphi$
applicable to non-relativistic shocks,
by a practical way.

First,
because the changes in $q$ and $t$ at one-cycle are expected to be mutually independent,
the dependences of $\varphi$ on $\Dq$ and $\Dt$ can be separable
as $\varphi(\Dq,\Dt) = \varphi_q(\Dq) \varphi_t(\Dt)$.
Since the particle distribution function at the shock front is approximately isotropic,
we have
\begin{equation}
	\Dqmean \sim \frac{4}{3} \frac{\Vu - \Vd}{c},
	\quad
	\sgmDq \sim \frac{1}{3} \frac{\Vu - \Vd}{c}.
	\label{eq:Dq_sgmDq_approx}
\end{equation}
[cf. equation \refp{eq:energy gain}, or equation (7) of \citealt{Bell78}].
These two quantities
are expected to be small enough,
compared with $\lambda^{-1}$,
to use
the approximations described
in Sections \ref{sec:approximation steady} and \ref{sec:approximation time dep}.
Therefore,
using these approximations,
we can proceed without specifying
the functional form of $\varphi_q$ here.

The time dependent part $\varphi_t$
can be approximated as follows
with the aid of a solution of the diffusion-convection equation
and results of Monte Carlo simulations.
In a one-dimensional diffusion process
with a diffusion coefficient $D$
in a moving medium with speed $V$,
the time distribution function
for returning from a distant point at $x=a$ ($a>0$)
to a barrier located at $x=0$
is given by
\begin{equation}
	g(t) = \frac{a}{\sqrt{4\pi D t^3}} \exp\squarebr{ -\frac{(a+Vt)^2}{4 D t} }
	\label{eq:gt}
\end{equation}
or in the form of the Laplace transform
\begin{equation}
	\tilde{g}(s) = \exp\left\{-\frac{a}{2D} \squarebr{V+(V^2+4Ds)^{1/2}} \right\}
\end{equation}
\citep[see][]{Cox,Lagage},
where $V$ can be positive or negative
and, for the present problem,
$V=-\Vu$ for the upstream region
and $V=\Vd$ for the downstream region.
Since
this function already
includes the influence of the escape of particles,
the return probability is given by
\begin{equation}
	\Pret(a) = \tilde{g}(0) =
		\left\{
			\begin{array}{ll}
				1 & \mbox{for $V < 0$} \\
				\exp(-aV/D)  & \mbox{for $V > 0$.}
			\end{array}
		\right.
		\label{eq:Pret_diffusion}
\end{equation}
First two moments of the return time $t$ are given by
\begin{equation}
	\anglebr{t} = \frac{a}{|V|},
	\qquad
	\anglebr{t^2}-\anglebr{t}^2 = \frac{2aD}{|V|^3}.
	\label{eq:gt moments}
\end{equation}

\begin{figure}
  \includegraphics[width=\linewidth]{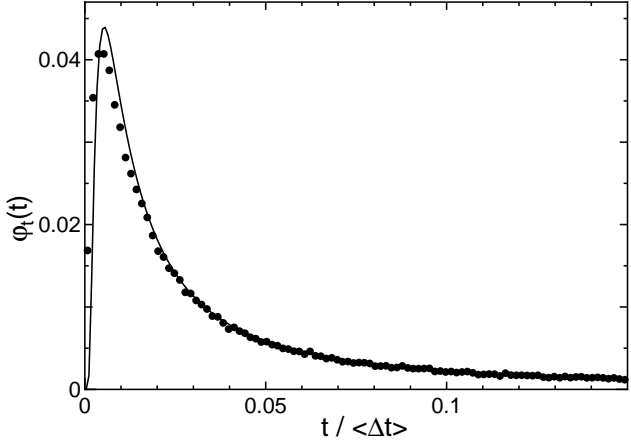}
  \caption{
  	Time distribution for one-cycle of particles
  	in a non-relativistic shock with $\Vs=0.01c$, $r=4$.
  	The solid curve is the approximate solution, $G(t)$,
  	obtained by inverting equation \refp{eq:tilG} numerically.
  	The dots are results from a Monte Carlo simulation.
  }
  \label{fig:cycle_time_dist}
\end{figure}

However,
the above results are not applicable directly
to calculate the cycle-time;
in order to do this,
we need the distribution of the residence time,
that is,
the time between
entering one of the fluid regions
and leaving the region through the shock front,
in each region.
Here,
we employ
Monte Carlo simulations
to obtain an approximate expression
for the residence time distribution.
It is known that,
in non-relativistic shocks,
the properties of particle motion
have little dependence
on the details of the scattering process.
Thus,
we can utilize the large-angle scattering model,
which was dealt with the previous paper \citep{Kato01},
in the simulation
to estimate the residence time
for isotropic injection of particles
at the shock front.
Performing the Monte Carlo simulations,
we found that
the distribution of the residence time
for isotropic injection at the boundary
\textit{is approximated well}
by $g(t)$ in equation \refp{eq:gt}
if we set the parameters for the diffusion process as follows:
\begin{equation}
	a = \frac{4}{3} c \tau_0,
	\qquad
	D = \frac{1}{3} c^2 \tau_0,
	\label{eq:a and D}
\end{equation}
where $\tau_0$ is the mean free time of the particles,
defined for the large-angle scattering model,
measured in the fluid rest frame.
Thus,
we can use $g(t)$ with $a = 4D/c$
as an approximation of the residence time distribution for each region.
Taking the convolution between
the residence time distributions for the upstream and downstream region,
we finally obtain
the distribution of the cycle-time
in the form of Laplace transform
\begin{eqnarray}
	\tilvphi_t(s) \sim \tilG(s)
	 = \exp\Bigl\{ -2\bigl[ \nud - \nuu
						+ (\nuu^2 + \frac{4\Du}{c^2} s)^{1/2} \nonumber \\
						+ (\nud^2 + \frac{4\Dd}{c^2} s)^{1/2} \bigr] \Bigr\},
		\label{eq:tilG}
\end{eqnarray}
%
where $\nuu := \Vu/c$ and $\nud := \Vd/c$;
the diffusion coefficients
in the upstream and downstream region
are denoted by $\Du$ and $\Dd$, respectively.
Fig.~\ref{fig:cycle_time_dist} shows the cycle-time distribution
calculated  for $\Vu=0.01c$ and $r=4$,
where we take $\Du=\Dd=D$.
The solid curve represents the result obtained by 
inverting equation \refp{eq:tilG} numerically
and
the dots are the results from the Monte Carlo simulation.
We see that the approximation
fits the simulation results quite well.
Note that,
this distribution is very skew and peaks at $t \ll \Dtmean$,
which are prominent features of the diffusion process
owing to the random walk motion of the particles.

Thus,
we can write an approximated form of $\tilvphi$:
\begin{equation}
	\tilvphi(\Dq,s) = \varphi_q(\Dq) \tilG(s)
	\label{eq:non-rela_tilvphi}
\end{equation}
with the quantities in \refp{eq:Dq_sgmDq_approx}.
Obviously,
the return probability at one-cycle is given by
\begin{equation}
	\Pret = \tilG(0) = \exp(-4\nud) \sim 1 - 4\nud,
\end{equation}
which is essentially in agreement with the previous results
\citep[e.g.][]{Bell78}.
The first two moments of the cycle-time are given by
\begin{equation}
	\Dtmean = \frac{4}{c} \left( \frac{\Du}{\Vu} + \frac{\Dd}{\Vd} \right),
	\qquad
	\sgmDt^2 = \frac{8}{c} \left( \frac{\Du^2}{\Vu^3} + \frac{\Dd^2}{\Vd^3} \right).
	\label{eq:t_cyc_mean}\\
\end{equation}
The former result coincides with the previous result
derived from the diffusion-convection equation \citep{Drury83}.
The state points shown in
Fig.~\ref{fig:sequence} were
also obtained by the Monte Carlo simulations performed above
with the same parameter as in Fig.~\ref{fig:cycle_time_dist}.

\subsection{Steady state and time-dependent solution}
Because of the isotropy of the particle distribution at the shock front,
we have
\begin{equation}
	\Fsfp(q,t) = \frac{1}{2} \Fsf(q,t),
	\qquad
	\mup = \frac{1}{2}.
\end{equation}
Because $\anglebr{\Dq} \ll 1$ and $\Pret \sim 1$,
using the approximations in (\ref{eq:lambda A approx2}),
we obtain
\begin{equation}
	\lambda \sim \frac{\Pesc}{\anglebr{\Dq}} \sim \frac{3}{r - 1},
	\qquad
	A \sim \anglebr{\Dq} \sim \frac{4}{3} \frac{\Vu-\Vd}{c}.
\end{equation}
Substituting the above results into equation (\ref{eq:fzp asymptotic}),
the steady state solution for large $p$
is given by
\begin{equation}
	\Fsf^{(p)}(p, \infty)
	= \frac{3Q_0}{\Vu-\Vd} \frac{1}{p_0}
				\left( \frac{p}{p_0} \right)^{-\sigma}
\end{equation}
with $\sigma = (r+2)/(r-1)$,
where $\Fsf^{(p)}(p) dp = \Fsf(q=\ln(p/p_0)) dq$.
In terms of the isotropic part of the usual phase-space distribution function $f(p,x,t)$,
where $x$ is distance from the shock front,
we have the following relation:
\begin{equation}
	f(p,0,t) = \frac{1}{4\pi p^2} \Fsf(p,t) = \frac{1}{4\pi p^3} \Fsf(q,t).
	\label{eq:f-F relation}
\end{equation}
We therefore obtain
\begin{equation}
	f(p,0,\infty) = \frac{3 Q_0'}{\Vu-\Vd} \frac{1}{p_0}
						\left( \frac{p}{p_0} \right)^{-(\sigma+2)},
	\label{eq:phase-space steady solution}
\end{equation}
where $Q_0'$ is
the mono-energetic, isotropic injection rate defined for $f(p,0,t)$,
with which $(df/dt)_\rmn{inj} = Q_0' \delta(p-p_0)$ at the shock front.
Note that,
$Q_0'$ has the following relation with $Q_0$:
\begin{equation}
	Q_0' = \frac{1}{4\pi p_0^2} Q_0.
	\label{eq:injection rate relation}
\end{equation}
The above result \refp{eq:phase-space steady solution}
is equivalent to the previous results
[e.g. equation (3.24) of \citealt{Drury83}
\footnote{The factor $1/p_0$ was however missing there.}
].

\begin{figure}
  \includegraphics[width=\linewidth]{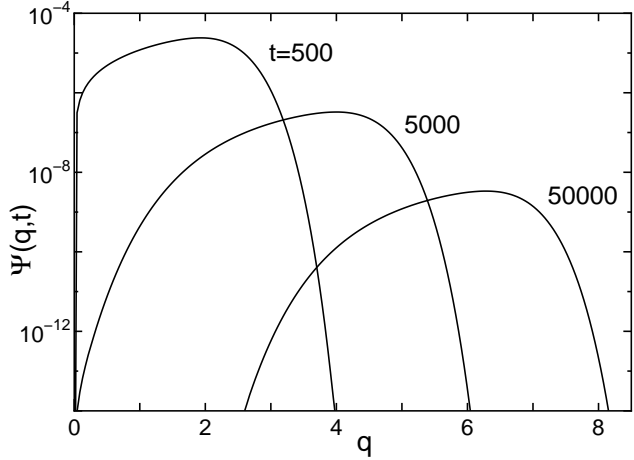}
  \caption{
  	Time development of
  	the density of state points $\Psi(q,t)$
  	for a non-relativistic shock with $\Vs=0.01c$ and $r=4$.
  	We take $\alpha=1$, $\Du=\Dd$ and
  	$t=500, 5000, 50000$, where the unit of time is 
  	the mean cycle-time for $q=0$.
  	The self-similarity given by equation \refp{eq:self-similar_Psi}
  	is evident for large $q$.
  }
  \label{fig:Psi}
\end{figure}

\begin{figure}
  \includegraphics[width=\linewidth]{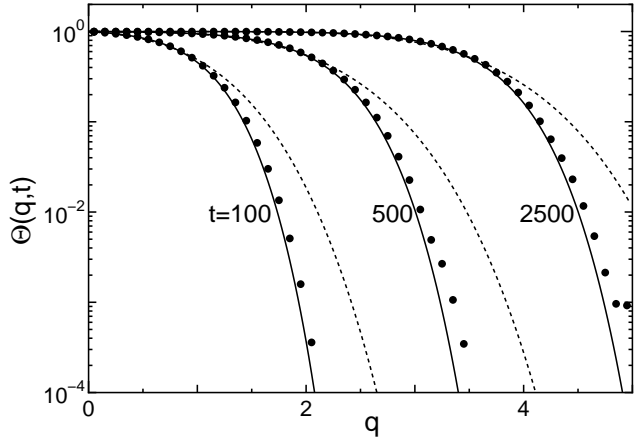}
  \caption{
  	Time development of
  	the cut-off function $\Theta(q,t)$.
  	The parameters are the same as in Fig.~\ref{fig:Psi},
  	except that t=100, 500, 2500.
  	The solid curves are numerically inverted solutions
  	by the present method,
  	and the dots are results from Monte Carlo simulations.
  	The dotted curves are
	the approximate solution of Drury (1991).
  }
  \label{fig:Theta}
\end{figure}

The time-dependent solution is obtained as follows.
Adopting the approximation \refp{eq:approx_tilPsi},
we can obtain the solution of $\tilPsi(q,s)$
with
\begin{equation}
	B_1(q,s) \sim \tilG(\erm^{\alpha q} s),
	\qquad
	B_2(q,s) \sim \Dqmean \tilG(\erm^{\alpha q} s),
\end{equation}
\begin{equation}
	I(q,s) = \frac{-q}{\Dqmean}
					+ \frac{1}{\Dqmean} \int_0^q \frac{1}{\tilG(\erm^{\alpha q'} s)} dq'.
\end{equation}
Then,
$\tilTheta(q,s)$ is given
by equation \refp{eq:tilTheta}.
Inverting these results numerically,
we finally obtain $\Psi(q,t)$ and $\Theta(q,t)$.
Fig.~\ref{fig:Psi} represents the time development of $\Psi(q,t)$
calculated for $\Vs=0.01c$, $r=4$, $\alpha=1$, $\Du=\Dd=D$
and $t=500, 5000, 50000$ in the unit of $\tcycz$.
The self-similarity
given by equation \refp{eq:self-similar_Psi} is evident.
Fig.~\ref{fig:Theta} represents the time development of $\Theta(q,t)$
calculated for the same parameters as in Fig.~\ref{fig:Psi}
except that $t=100, 500, 2500$.
The solutions given by the present method (solid curves)
fit the results from Monte Carlo simulations (dots)
quite well.

It is interesting
to make a comparison between
the cut-off function by our method
and that by the approximation of \cite*{Drury91}.
The first two cumulants $c_1, c_2$
in his approximation
[see equations (25) and (26) in his paper]
are determined as follows.
In the present situation,
the diffusion coefficients are expressed as
$\Du = \Dd = D_0 p/p_0$,
where $D_0$ is the diffusion coefficient for
particles with the injection momentum $p_0$.
From equation \refp{eq:t_cyc_mean}
we can determine $D_0$
in terms of $\tcycz$ as
\begin{equation}
	D_0 = \frac{c}{4} \left( \frac{1}{\Vu} + \frac{1}{\Vd} \right)^{-1} \tcycz.
\end{equation}
Using this relation,
we determine for the present situation
\begin{equation}
	c_1 = \frac{3c}{4\Vs} \frac{r}{r-1} \left( p/p_0 - 1 \right) \tcycz,
\end{equation}
\begin{equation}
	c_2 = \frac{3c^2}{16\Vs^2} \frac{r(r^3+1)}{(r-1)(r+1)^2}
		\left[ (p/p_0)^2 - 1 \right] \tcycz^2.
\end{equation}
Substituting these results into
equation (22) of \cite*{Drury91},
we obtain the approximated cut-off function
represented in Fig.~\ref{fig:Theta}
by dotted curves.
As mentioned in his paper,
this approximation
do not provide satisfactory fits
for high energy tail of the distributions,
while it provides fairly good fits for lower energies
[see also fig.~4 of his paper].

\section{Relation between the one-cycle properties and particle distribution}
\label{sec:Discussion}
Recently,
the acceleration of particles
in ultrarelativistic shocks
has attracted some attention in relation to
the ultrahigh-energy cosmic rays
or high-energy particles at the external shocks associated with gamma-ray bursts.
Many works on this issue
(theories and simulations) showed that
in such shocks
the energy gain at one-cycle is approximately given by $\Dq \sim \ln 2$,
owing to
the characteristics of the trajectory of particles in the upstream region,
and the power-law index is typically given by $\sigma \sim 2.2$
\citep[e.g.][]{Bednarz98,Kirk00,Achterberg01}.
The deviation of $\Dq$, $\sgmDq$,
becomes the same order of $\Dqmean$.
Although
Peacock's formula \refp{eq:lambda approx1}
has been often used
in the literature
to calculate the power-law index
\citep[e.g.][]{Kato01,Achterberg01},
it can deviate from the true value
because
the criterion \refp{eq:approx_criterion}
may not hold
in this situation.
In addition,
it has been shown that
the cycle-time can be dominated by
the upstream residence time \citep{Gallant99,Achterberg01}.
Because
the distribution of the upstream residence time
in ultrarelativistic shocks
can be much different from that of non-relativistic shocks
(cf.  Fig.~\ref{fig:cycle_time_dist}
and fig.10 of \citealt{Achterberg01}),
the feature of the cut-off function
can be also different from that of non-relativistic shocks considerably.

\begin{figure}
  \includegraphics[width=\linewidth]{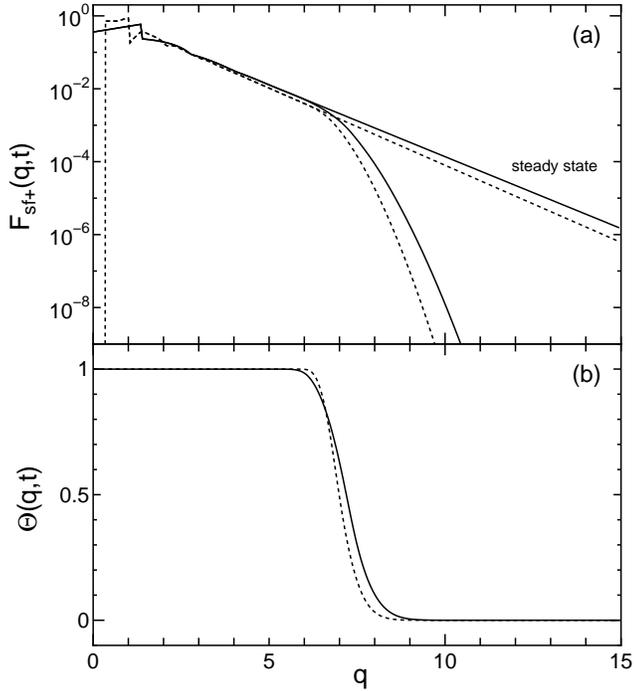}
  \caption{
  	The distribution function of particles
  	at the shock front
  	for steady injection (a),
  	and the cut-off function (b).
  	The time-dependent solutions are calculated at $t=10^3$.
  	The solid curves denote model 1 and the dotted curves denote model 2.
  }
  \label{fig:model1_2}
\end{figure}

To give some insights into these issues,
we briefly investigate here
how the properties of the one-step probability density function,
that is, the stochastic properties of $\Dq$ and $\Dt$,
affect the shape of the resulting particle distribution,
for several toy models.
We assume the scaling law \refp{eq:rho_sym}
discussed in Section \ref{sec:Solution}
and the following separable form
\begin{equation}
	\varphi(\Dq,\Dt) = \varphi_q(\Dq) \varphi_t(\Dt).
\end{equation}
For the following all models,
we adopt $\Pret=0.5$, $\Dqmean=\ln 2$, $\tcycz=1$ and $\alpha=1$.
First,
in order to investigate
the influence of the dispersion of $\Dq$,
we consider the following two simple models:
\begin{equation}
	\varphi_q(\Dq) = \dfrac{H(\Dq)-H(\Dq-2\Dqmean)}{2\Dqmean}
\end{equation}
(model 1), and
\begin{equation}
	\varphi_q(\Dq) = \dfrac{H(\Dq-\Dqmean/2)-H(\Dq-3\Dqmean/2)}{\Dqmean}
\end{equation}
(model 2),
where $H(x)$ is the unit step function:
$H(x)=0$ for $x<0$, $H(x)=1$ for $x>0$
and $H(x)=1/2$ for $x=0$.
The standard deviation of $\Dq$ for model 2 is a half of that for model 1.
For both models,
we adopt
\begin{equation}
	\varphi_t(\Dt) = \Pret \dfrac{H(\Dt)-H(\Dt-2\Dtmean)}{2\Dtmean}.
	\label{eq:vphi_t_1}
\end{equation}
These models may represent typical features of
the energy-gain and cycle-time in ultrarelativistic shocks
which were examined by \citet{Achterberg01} by Monte Carlo simulations.
For model 1,
solving equation \refp{eq:lambda determine} numerically,
we obtain the power-law index of $\lambda \sim 0.91$,
while model 2 gives $\lambda \sim 0.98$.
This result explicitly shows that
the power-law index $\lambda$ is generally dependent
not only on $\Dqmean$ but also on the dispersion $\sgmDq$,
although this fact is rather evident from equation \refp{eq:lambda determine}.
Because
the criterion \refp{eq:approx_criterion}
is not satisfied for both models,
especially for model 1,
Peacock's formula \refp{eq:lambda approx1},
which gives $\lambda = 1$ for both models,
deviates from the true value.
[the dispersion of $\Dq$
is neglected in the derivation of equation \refp{eq:lambda approx1}
and that of \citet{Peacock81}.]
This fact may also be responsible for
the small discrepancy in the power-law indices
in fig.14 of the previous paper \citep{Kato01}
between our results, which utilized the formula \refp{eq:lambda approx1},
and those of \cite*{Ellison90},
which were determined by fitting results from Monte Carlo simulations.
Fig.~\ref{fig:model1_2} (a) shows
the steady state and time-dependent solutions
of the distribution function of particles
(in an arbitrary unit)
and Fig.~\ref{fig:model1_2} (b) shows the cut-off functions,
for both models.
These solutions are obtained by the numerical method
given in Section \ref{sec:Numerical Solution}.
The time-dependent solutions are calculated at $t=10^3$.
It is seen that
model 2 (dotted curve) has a sharper cut-off than model 1 (solid curve),
owing to the smaller dispersion of $\Dq$ than that of model 1.

\begin{figure}
  \includegraphics[width=\linewidth]{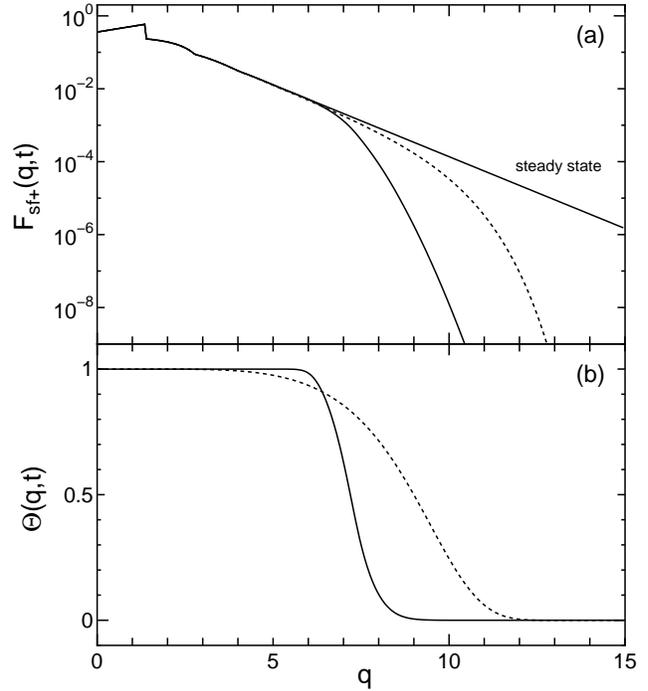}
 	\caption{
 		The same as Fig.~\ref{fig:model1_2}
 		except that the dotted curves denote model 3.
 		Although both models have the same mean cycle-time,
 		the shape of the high energy tails
 		are considerably different from each other.
  }
  \label{fig:model1_3}
\end{figure}

Then,
we investigate the influence of the dispersion of $\Dt$
on the distribution function.
We define model 3,
which has a `diffusive' nature
on the cycle-time, by
\begin{equation}
	\varphi_t(\Dt) = \Pret g(\Dt)/\exp(-4V/c),
\end{equation}
where $g(t)$ was defined in equation \refp{eq:gt}.
Although
$g(t)$ was derived
from the diffusion-convection equation
for the return-time distribution function,
not for the cycle-time,
it may partly represent essential behaviour
of the cycle-time distribution
for the present situation
when the motion of particles is mainly diffusive.
In this equation,
we use the \textit{normalized} distribution $g(\Dt)/\exp(-4V/c)$,
where $V=0.01c$, $a=V\tcycz$ and $D=cV\tcycz/4$
so that $\tcycz = 1$
[cf. equations \refp{eq:Pret_diffusion}--\refp{eq:a and D}],
and take the return probability $\Pret=0.5$ independently.
$\varphi_q(\Dq)$ is taken the same as in model 1;
the spectral index $\lambda$ of model 3
therefore equals to that of model 1.
Fig.~\ref{fig:model1_3} (a) shows
the distribution function of particles
and (b) the cut-off function,
for model 1 (solid curve) and model 3 (dotted curve)
as in Fig.~\ref{fig:model1_2}.
Although both models have the same mean cycle-time $\Dtmean$,
the shapes of high-energy tails are considerably different.
This is attributed to the large dispersion of $\Dt$ in model 3.

\begin{figure}
  \includegraphics[width=\linewidth]{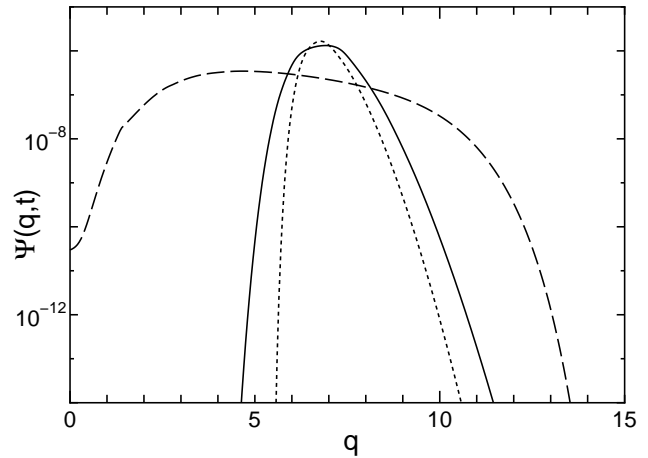}
 	\caption{
   	The density of state points for 
   	model 1 (solid curve), model 2 (dotted curve) and model 3 (dashed curve).
   	These are calculated at $t=10^3$.
  }
  \label{fig:Psi_123}
\end{figure}

These features
may be observed from
the densities of state points
shown for the three models
in Fig.~\ref{fig:Psi_123}.
Note that
all models have the same mean cycle-time $\Dtmean$
and
these densities are proportional to the distribution functions of particles
at the shock front for an impulsive injection.
In particular for the diffusive model (model 3),
on account of its large dispersion of $\Dt$,
there are particles which are accelerated effectively
compared with the other models
while inefficiently accelerated particles also exist.

\section{Conclusions}
\label{sec:Conclusions}
In this paper,
we have given a new description of
the first-order Fermi acceleration
in shock waves
based on a stochastic, single-particle approach.
The method can treat
the time development of
the particle distribution
and determine the power-law index exactly,
even for relativistic shocks.

In Section \ref{sec:Description},
we have formulated the mechanism
from the point of view of stochastic process.
Representing the properties of the one-cycle of shock crossings
of particles
by a probability density function $\rho$,
we have formulated the acceleration process
as a two-dimensional Markov process
on the $q-t$ plane.
Introducing the density of state points in equation \refp{eq:Psi_sum},
we have derived a fundamental integral equation \refp{eq:Integral eq}
for describing this stochastic process.
Then
we have related the density of state points
with the distribution function of particles at the shock front.
We have also given numerical methods to solve the integral equation.

In Section \ref{sec:Solution},
we have investigated
the case where
the one-step probability density function satisfies the scaling law
given by equation \refp{eq:rho_sym}.
Investigating the asymptotic behaviour
of the steady state solution,
we have confirmed that
these are characterized by the power-law distribution in momentum space,
and derived the equation \refp{eq:lambda determine}
which determines the power-law index exactly
for general cases.
We have also shown that
the time-dependent solutions satisfy
the similarity law given by equation \refp{eq:self-similar_Psi}
except for momentum near the injection.
Then
we have derived an approximate solution
which is applicable to the acceleration in non-relativistic shocks.

In Section \ref{sec:Application},
we have applied the theory
to non-relativistic shocks
as a check of our method.
We have confirmed that the steady state solution
obtained by our method
coincides with the well-known previous results
obtained by various methods
and that
the time-dependent solution
essentially coincides with the results of Monte Carlo simulations.

Finally,
we have examined
the relation between the property of the one-step probability density function
and the steady and time-dependent solution,
in Section \ref{sec:Discussion}.
We have explicitly shown
that
the power-law index $\lambda$ generally depends
not only on the mean energy gain $\Dqmean$
but also on the dispersion of $\Dq$, $\sgmDq$.
This can be important for relativistic shocks.
We have shown
that the distribution of $\Dt$
for the diffusion process
results in a quite broad high-energy tail
in the distribution function of particles and the cut-off function
compared with the uniform distribution such as \refp{eq:vphi_t_1}
even for the same mean cycle-time,
because of its large dispersion $\sgmDt$.

\section*{Acknowledgments}
This work is supported in part by ACT-JST
\footnote{
Research and Development for Applying Advanced Computational Science 
and Technology of Japan Science and Technology Corporation
}
(T.N.K.)
and
in part by
a Grant-in-Aid for Scientific Research
from the Ministry of Education and Science of Japan
(No.13440061, F.T.).

\appendix

\section{Numerical inversion of Laplace transforms}
\label{sec:Inversion}
Here,
we briefly explain
the numerical method for inversion of Laplace transforms
used in this paper.
The method is based on 
\citet{Abate} \citep[see also][]{Hosono}.
First,
the inversion formula of Laplace transform is given by
\begin{eqnarray}
	f(t)
	&=& \frac{1}{2 \pi \irm}
				\int_{\gamma - \irm\infty}^{\gamma + \irm\infty} \tilde{f}(s) \erm^{s t} ds
					\nonumber \\
	&=& \frac{2 \erm^{\gamma t}}{\pi}
				\int_0^\infty \rmn{Re} \left[ \tilde{f}(\gamma+\irm y) \right] \cos(yt) dy,
\end{eqnarray}
where $\gamma$ is a real constant taken so that
the vertical line $\rmn{Re}(s)=\gamma$ lies on the right of
the all poles of $\tilde{f}(s)$.
Applying the trapezoidal rule to calculate the above integral
with the step size $\Delta y = \pi/(2t)$,
we obtain
\begin{equation}
	f(t) \sim \frac{\erm^{A/2}}{t} \left[ \frac{1}{2} a_0(t) + \sum_{m=1}^n (-1)^m a_m(t) \right],
	\label{eq:inversion formula}
\end{equation}
where
$A$ relates to the numerical precision
($A=\eta \ln 10$ for the precision of $10^{-\eta}$),
and
\begin{equation}
	a_m(t) := \rmn{Re} \left[ \tilde{f} ( X + \irm mH ) \right]
\end{equation}
with $X:=A/(2t)$ and  $H:=\pi/t$.
The number of terms to be summed, $n$,
should be chosen carefully for each problem.
(In this paper, we choose $n=80$.)
In order to calculate
the sum in equation (\ref{eq:inversion formula}),
which is an alternating series if $a_m$ does not change the sign,
we apply Euler's transformation with van Wijgaarden's algorithm (see chapter 5.1 of \citealt{NR}).
The essence of
our code implemented in C++ is as follows:
\begin{verbatim}
#include  <complex>
using namespace std;
typedef complex<double> cplxd;
const double PI = 3.141592653589793238;

double
invLaplace(cplxd Fs(const cplxd&),
           double t, int n, double A)  {
  const int	nmax = 500;
  const double X(0.5*A/t), H(PI/t);
  if (n > nmax)	n = nmax;

  // Euler summation (van Wijngaarden's algorithm)
  double sum, sgn, wk[nmax+1];
  int k = 0;
  sgn = -1.0;
  sum = 0.5*(wk[0]=0.5*real(Fs(cplxd(X,0))));
  for (int m=1; m<n; m++, sgn=-sgn) {
    double nxt, cur;
    nxt = sgn*real(Fs(cplxd(X,m*H)));
    for (int j=0; j<=k; j++) {
      cur = wk[j]; wk[j] = nxt;
      nxt = 0.5*(cur+nxt);
    }
    sum += (fabs(nxt)>fabs(cur)) ?
              nxt : 0.5*(wk[++k]=nxt);
  }
  return exp(0.5*A)/t * sum;
}
\end{verbatim}
Here,
we use the template class \verb@complex@
included in the C++ standard library.
In the arguments of the function,
\verb@Fs(const cplxd& s)@ is a user-supplied, complex-valued function $\tilde{f}(s)$ to be inverted,
\verb@t@ is the time $t$ for evaluation,
\verb@n@ is the number of terms to be summed in \refp{eq:inversion formula} $n$,
and
\verb@A@ is $A$ mentioned above.


\end{document}